\documentclass[usenatbib]{mn2e}
\usepackage{psfig}
\usepackage[colour]{optional}

\newcommand{\Msun}{\mbox{$\rm M_{\odot}$}} 
 
\newcommand{\Zsun}{\mbox{$\rm Z_{\odot}$}}
\newcommand{\LCDM}{\mbox{$\Lambda$CDM}}

\newcommand{\Tvir}{\mbox{$T_{\rm vir}$}}

\newcommand{\htwo}{\mbox{H$_2$}} 
\newcommand{\htwosub}{\mbox{\scriptsize H$_2$}} 
\newcommand{\e}{\mbox{e$^-$}}
\newcommand{\h}{\mbox{H}} 
\newcommand{\hplus}{\mbox{H$^+$}}
\newcommand{\htwoplus}{\mbox{H$_2^+$}} 
\newcommand{\hminus}{\mbox{H$^-$}}
\newcommand{\he}{\mbox{He}} 
\newcommand{\heplus}{\mbox{He$^+$}}
\newcommand{\heplussub}{\mbox{\scriptsize He$^+$}}
\newcommand{\heplusplus}{\mbox{He$^{++}$}}

\newcommand{\spose}[1]{{\hbox to 0pt{#1\hss}}}
\newcommand{\lta}{\mathrel{\spose{\lower 3pt\hbox{$\mathchar"218$}}
     \raise 2.0pt\hbox{$\mathchar"13C$}}}
\newcommand{\gta}{\mathrel{\spose{\lower 3pt\hbox{$\mathchar"218$}}
     \raise 2.0pt\hbox{$\mathchar"13E$}}}
\newcommand{\approxlt}{\mathrel{\spose{\lower 3pt\hbox{$\sim$}}
	\raise 2.0pt\hbox{$<$}}}
\newcommand{\approxgt}{\mathrel{\spose{\lower 3pt\hbox{$\sim$}}
	\raise 2.0pt\hbox{$>$}}}
\newcommand{\approxpropto}{\mathrel{\spose{\lower 3pt\hbox{$\sim$}}
	\raise 2.0pt\hbox{$\propto$}}}
\newcommand{\twiddles}[1]{\spose{\raise 5pt\hbox{$\sim$}}\hbox{$#1$}}

\title
  [Primordial star clusters under UVB radiation]
  {On the population of primordial star clusters in the presence of UV
  background radiation}

\author
  [M. A.  MacIntyre, F. Santoro \& P. A. Thomas]
  {Michael A.~MacIntyre$^1$\thanks{E-mail: m.a.macintyre@sussex.ac.uk},
   Fernando Santoro$^2$ and Peter A.~Thomas$^1$\\
   {}$^1$Astronomy Centre, University of Sussex, Falmer, Brighton,
   BN1\,9QH, UK\\
   {}$^2$Center for Astrophysics and Space Astronomy, University of
   Colorado, 389 UCB, Boulder, Colorado 80309-0389, USA
}

\date{\today}

\begin{document}
\journal{Preprint astro-ph/0510074}
 
\maketitle

\begin{abstract}

We use the algorithm of \cite{CLBF00} to generate merger trees for the
first star clusters in a $\Lambda$CDM cosmology under an isotropic UV
background radiation field, parametrized by $J_{21}$. We have
investigated the problem in two ways: a global radiation background
and local radiative feedback surrounding the first star clusters.

Cooling in the first halos at high redshift is dominated by molecular
hydrogen, \htwo---we call these Generation~1 objects.  At lower
redshift and higher virial temperature, $T_{\rm vir}\gta10^4$K,
electron cooling dominates---we call these Generation~2.  

Radiation fields act to photo-dissociate \htwo, but also generate free
electrons that can help to catalyse its production.  At modest
radiation levels, $J_{21}/(1+z)^3\sim10^{-12}-10^{-7}$, 
the nett effect is to enhance the formation of Generation~1
star-clusters.  At higher fluxes the heating from photo-ionisation
dominates and halts their production.  With a realistic build-up of
flux over time, the period of enhanced \htwo\ cooling is so fleeting
as to be barely discernable and the nett effect is to move primordial
star cluster formation towards Generation~2 objects at lower redshift.

A similar effect is seen with local feedback.  Provided that enough
photons are produced to maintain ionization of their host halo,
they will suppress the cooling in Generation~1 halos and boost the 
numbers of primordial star clusters in Generation~2 halos.  Significant
suppression of Generation~1 halos occurs for specific photon fluxes
in excess of about $10^{43}$\,ph\,s$^{-1}\,\Msun^{-1}$.

\end{abstract}


\section{Introduction}
\label{sec:intro}

Primordial star clusters contain the first stars to form in the
Universe, from zero-metallicity gas.  Previous work
(e.g.~\citealt{TSR97}; \citealt{AAN98}; \citealt{HST02};
\citealt{BCL99}; \citealt{ST03}) has concentrated on the very first
objects for which there is no significant external radiation
field. However, these first clusters are expected to produce massive
stars which will irradiate the surrounding Universe and may well be
responsible for partial re-ionisation of the intergalactic medium. 

The only species produced in sufficient abundance to affect the
cooling at early times is molecular hydrogen. Its presence allows the
first objects to cool and form in low temperature halos ($T<10^4K$) at
high redshift ($z_{\rm vir} \sim 20-30$). However, molecular hydrogen
is very fragile and can easily be dissociated by UV photons in the
Lyman-Werner bands, ($11.2-13.6\rm{eV}$). Thus, the formation of the first
stars may well have a negative feedback effect on subsequent
population \rm{III} star formation by supressing cooling via this
mechanism. This problem is not a trivial one and has been the subject of
much interest in recent years (e.g.~\citealt{HRL96}; \citealt{HAR00};
\citealt{KTUSI00}; \citealt{GB01}; \citealt{Omu01}; \citealt{MBA01};
\citealt{RGS02}; \citealt{OH02}; \citealt{Cen03}; \citealt{CFW03};
\citealt{YAHS03}, \citealt{TVS04}). The complexity of the feedback and
the large number of unknowns (e.g. population \rm{III} IMF, total
ionising photon production, etc.) make this problem very  challenging.    

In an attempt to understand this era of primordial star cluster
formation, we investigated in a previous paper \citep[hereafter ST03]
{ST03} the merger history of primordial haloes in the
\LCDM~cosmology. There we assumed no external radiation field (other
than that provided by CMB photons).  The Block Model of \cite{CoK88}
was used to generate the merger history of star clusters using a
simple model for the collapse and cooling criterion, hence identifying
those haloes that were able to form stars before being disrupted by
mergers. We then contrasted the mass functions of all the resulting
star clusters and those of primordial composition, i.e.  star clusters
that have not been contaminated by sub-clusters inside them.  We found
two generations of primordial haloes: low-temperature clusters that
cool via \htwo, and high-temperature clusters that cool via electronic
transitions.

We investigated two regions of space each enclosing a mass of
$10^{11}h^{-1}\Msun$: a high-density region corresponding to a
$3\sigma$~fluctuation ($\delta_0=10.98$), and a mean-density region
($\delta_0=0$), where $\delta_0$ is the initial overdensity of the
root block. In the high-density region we found that approximately half
of the star clusters are primordial. The fractional mass contained in
the two generations was 0.109 in low temperature clusters and 0.049 in
high temperature clusters.  About 16 per cent of all baryons in this
region of space were once part of a primordial star cluster.  In the
low-density case the fractional mass in the two generations was almost
unchanged, but the haloes collapsed at much lower redshifts and the
mass function was shifted toward higher masses.

In the present paper, as a continuation of the previous work, we
include the effect of ionising radiation in two different ways:
firstly we add a homogeneous background radiation field; secondly, we
consider feedback from the first star clusters formed in the merger
tree---these will form an ionising (and photo-dissociating)
sphere around them, changing the cooling properties of neighbouring
star clusters. A further improvement upon our previous work includes
the use of a more realistic merger tree.

We describe our chemical network including radiative processes in
Section~\ref{sec:chemistry}, and the new merger tree method in
Section~\ref{sec:tree}.   The effect of a global ionisation field on
the formation of stars in primordial star-clusters is considered in
Section~\ref{sec:global} and that of local feedback in
Section~\ref{sec:local}.  Finally, we summarise our conclusions in
Section~\ref{sec:raddiscuss}.

\section{Primordial chemistry and gas cooling in the
presence of radiation}
\label{sec:chemistry}

\subsection{Chemical model}
\label{sec:chemmodel}

In this Section we introduce the chemical network needed to follow the
coupled chemical and thermal evolution under a homogeneous UV
background radiation field.

The non-equilibrium chemistry code is based upon the minimal model
presented in \citet[hereafter HSTC02]{HST02}. It calculates the
evolution of the following 9 species: \htwo, \h, \hplus, \htwoplus,
\hminus, \he, \heplus, \heplusplus and \e.  The important cooling
processes are: molecular hydrogen cooling, collisional excitation and
ionisation of atomic Hydrogen, collisional excitation of \heplus, and
inverse Compton cooling from cosmic microwave background photons. In
this paper we only consider the low density-high temperature ($\rm T >
300K$) limit. Thus, we have ignored the effects of HD cooling, which
is only important in the high density-low temperature regime.

To this chemical model we have added the photo-ionisation and
photo-dissociation reactions compile by \cite{AAZN97} and listed in
Table~\ref{radtab}.  This consists of 9 reactions involving the
interaction of each species with the background photons.  These are:
photo-ionisation of \h, \he, \heplus~and \htwo, with threshold
energies of $h\nu=$13.6, 24.6, 54.4 and 15.42\,eV, 
respectively; photo-detachment of \hminus with a threshold energy of
0.755\,eV, potentially an important process since \hminus catalyses
the formation of \htwo;  and photo-dissociation of \htwoplus~and
\htwo~(by the Solomon process and by direct photo-dissociation).  In
the case of the Solomon process dissociation happens in a very narrow
energy range $12.24$eV$<h\nu<13.51$eV.   

\begin{table*}
\caption{This table summarises the important reactions that should be
  included in a chemical network if a uniform background radiation
  field is present.  Compiled by Abel et al.~(1997), except cross
  sections 25 and 27.  The number index of each reaction corresponds to
  those in that paper.   Reference: Osterbrock (1989, O89), de Jong
  (1972, DeJ72), O'Neil \& Reinhardt (1978, OR78), Tegmark et al.~(1997,
  TSR97), Haiman, Rees \& Loeb (1996, HRL96), Abel et al.~(1997, AAZN97).}
  \nocite{O89}      \nocite{D72}      \nocite{OR78}     \nocite{TSR97}
  \nocite{HRL96} \nocite{AAZN97}
\label{radtab}
\begin{center}
\begin{tabular}{llll}
\emph{} & Reaction & cross sections/cm$^2$ & Reference \\[2pt] \hline

20 & $\h + \gamma \longmapsto \hplus +   2\e$    &
   $\sigma_{20}=A_0\left(\nu\over\nu_{th}\right)^{-4}~\left(e^{(4-4\arctan\epsilon/\epsilon)}\over(1-e^{-2\pi/\epsilon})\right)$~~$\left\{
   \begin{array}{r@{\quad\quad}l}     A_0=6.30\times10^{-18}    cm^2\\
   \epsilon= \sqrt[]{\nu/\nu_{th}-1}, ~~h\nu_{th}=13.6eV
     \end{array}\right.$
   & O89\\[2pt]

21    &   $\he +    \gamma \longmapsto \heplus +   \e$    &
   $\sigma_{21}=7.42\times10^{-18}\left(1.66\left(\nu\over\nu_{th}\right)^{-2.05}
   -0.66\left(\nu\over\nu_{th}\right)^{-3.05}\right), ~~\nu>\nu_{th}$ &
   O89 \\[2pt]

22   &   $\heplus +   \gamma \longmapsto \heplusplus +  \e$   &
   $\sigma_{22}=(A_0/Z^2)\left(\nu\over\nu_{th}\right)^{-4}~\left(e^{(4-4\arctan\epsilon/\epsilon)}\over(1-e^{-2\pi/\epsilon})\right)$, ~~
   $h\nu_{th}=Z^2\times13.6eV$~and $Z=2$.  & O89 \\[2pt]

23    &    $\hminus +    \gamma \longmapsto \h +   \e$    &
   $\sigma_{23}=7.928\times10^5(\nu-\nu_{th})^{3\over2}\left(1\over\nu^3\right), ~~h\nu>h\nu_{th}=0.755eV$
   & DeJ72 \\[2pt]

24   &    $\htwo +\gamma \longmapsto \htwoplus +    \e $   &
   $\sigma_{24}=\left\{     \begin{array}{r@{\quad:\quad}l}     0    &
   h\nu<15.42eV \\   6.2\times10^{-18}h\nu -9.40\times10^{-17}   &
   15.42<h\nu<16.50eV\\  1.4\times10^{-18}h\nu -1.48\times10^{-17}  &
   16.5<h\nu<17.7eV\\ 2.5\times10^{-14}(h\nu)^{-2.71} & h\nu>17.7eV
     \end{array}\right.$
   & OR78 \\[2pt]

25 & $\htwoplus + \gamma \longmapsto \h + \hplus$
   &    $\sigma_{25}=7.401\times10^{-18}10^{(-x^2    -   .0302x^3    -
   .0158x^4)}$~~$\left\{
      \begin{array}{r@{\quad\quad}l} 
	x=2.762\ln(h\nu/11.05eV)\\ h\nu>2.65eV
     \end{array}\right.$
   & TSR97 \\[2pt]

26   &   $\htwoplus +   \gamma \longmapsto 2\hplus +   \e$   &
   $\sigma_{26}=10^{-16.926                     -4.528\times10^{-2}h\nu
   +2.238\times10^{-4}(h\nu)^2
   +4.245\times10^{-7}(h\nu)^3}$, ~~30eV$<h\nu<$90eV & AAZN97 \\[2pt]

27 & $\htwo + \gamma \longmapsto \htwo^* \longmapsto \h + \h$
   &   $\sigma_{27}=3.71\times10^{-18}$, ~~12.24eV$<h\nu<$13.51eV &
   HRL96\\[2pt]

28 &  $\htwo +  \gamma \longmapsto \h + \h$  & see Reference for the
   expression & AAZN97 \\[2pt]

\end{tabular}
\end{center}
\end{table*}

The energy equation takes the following form:
\begin{equation}
\frac{d(n_{t}T)}{dt} = \frac{2}{3k} \left(\Lambda_{\rm{heat}}-
\Lambda_{\rm{cool}}\right)
\label{eq:enerequ2}
\end{equation}
where $n_t$ is the total number density of all species, $T$ is the
temperature, $\Lambda_{\rm{cool}}$~and $\Lambda_{\rm{heat}}$ are the
cooling and heating terms, respectively, and we have assumed a
monatomic energy budget of ${3\over2}kT$ per particle (the energy
associated with rotational and vibrational states of \htwo\ is
negligible).
\begin{equation}
\Lambda_{\rm cool} = \Lambda_{\rm H, ce} + \Lambda_{\rm H, ci} +
\Lambda_{\heplussub, {\rm ce}} + \Lambda_{\htwosub, {\rm ce}} + 
\Lambda_{\rm Compton}
\end{equation}
where the suffixes ce and ci mean collisional excitation and
ionisation respectively, and expressions for each of these terms are
given in HSTC02.
\begin{equation}
\Lambda_{\rm{heat}}=\Lambda_{\rm H, pi} + \Lambda_{\rm He, pi} + 
\Lambda_{\heplussub, {\rm pi}} + \Lambda_{\htwosub, {\rm pi}} + 
\Lambda_{\htwosub, {\rm pd}}
\end{equation}
where the suffixes pi and pd mean heating from photo-ionisation and
photo-dissociation respectively.  The expression for each term was
calculated using Equation \ref{eq:heatingt} from the Appendix, using
the cross-sections listed in Table \ref{radtab}.

\subsection{UVB spectrum}
\label{sec:uvb}
The non-equilibrium chemistry and the thermal evolution of the clouds
are calculated assuming the presence of an ultraviolet radiation field
of the power-law form
\begin{equation}
J_\nu=J_{21}\times\left(\nu\over\nu_H\right)^{-\alpha}10^{-21}\,
{\rm erg~s}^{-1}{\rm cm}^{-2}{\rm Hz}^{-1}{\rm sr}^{-1},
\label{eq:spectrum}
\end{equation}
where $h\nu_{\rm{H}}=13.6$\,eV is the Lyman limit of \h.  Here the
direction of the radiation field is not important and the
normalisation is given in terms of the equivalent isotropic field.

In the present-day Universe, QSOs have much steeper spectra
($\alpha\approx1.8$, e.g.~\citealt{ZKT97}) than do stars
($\alpha\approx5$, e.g.~\citealt{BL99}). However, Population~{\sc iii}
stars are likely to be biased to much higher spectral energies and their
spectra may resemble those of QSOs above the Lyman limit
(\citealt{TS00}; \citealt*{TSV03}).  In this paper, we take $\alpha=2$
which could equally well apply to either type of source.

For the calculation of all the photo-ionisation and photo-dissociation
rates and as well as the heating terms, we assume an optical
thin medium and no self-shielding.

\subsection{Cooling of isolated clouds}
\label{sec:isocool}

Apart from a scaling factor, the cooling time, $t_{\rm cool}$, of
isolated halos subject to a uniform radiation field depends solely on
the ratio of the number densities of baryons, $n_{\rm b}$, and
photons, $n_\gamma$, i.e.~$t_{\rm cool}=$fn$(T_{\rm vir},n_{\rm
  b}/n_\gamma)/n_{\rm b}$.  Thus the effect of the global background
radiation on single haloes can be presented in two ways: we can fix
either the baryon density or the amplitude of the radiation field and vary
the other.

\begin{figure}
\begin{center}
\opt{colour}{\psfig{width=9cm,angle=90,file=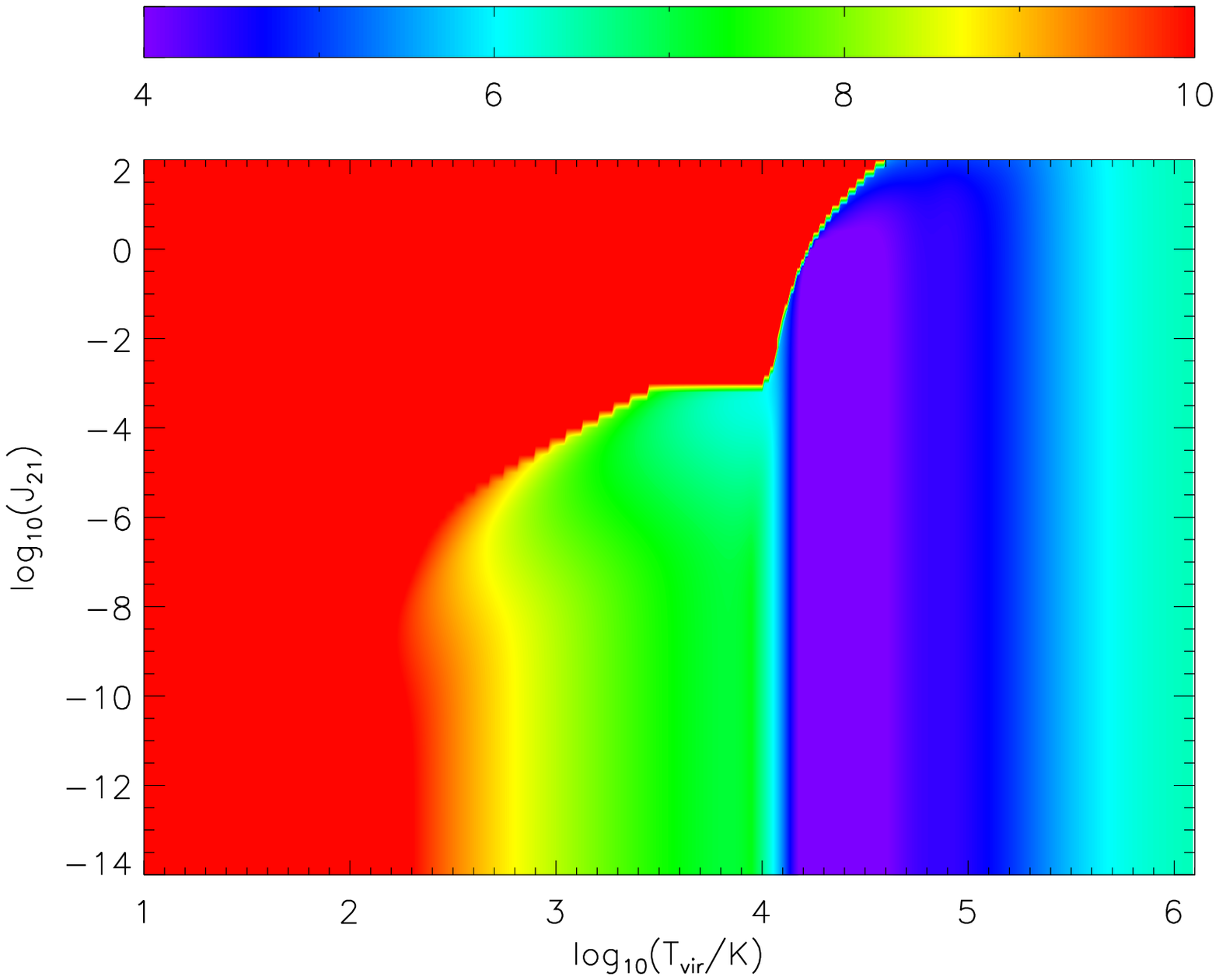}}
\opt{bw}{\psfig{width=9cm,angle=90,file=figure1_bw.ps}}
\caption{The cooling time, $t_{\rm cool}$ in the $J_{21}-T_{\rm{vir}}$ plane.
 The colour bar is $\rm{log}10(t_{\rm{cool}}/\rm{yr})$.  This
 plot was calculated at a fixed density corresponding to $z=20$.
 \opt{bw}{Note that this figure and figure~\ref{fig:fh2contours} look
 much nicer in colour in the on-line version of this paper.}}
\label{fig:3dcoolt}
\end{center}
\end{figure}

Figure \ref{fig:3dcoolt} shows the cooling time, $t_{\rm cool}$, of
haloes exposed to different levels of background radiation for a fixed
density of $n_H\approx0.31$\,cm$^{-3}$, corresponding to the mean
density within a collapsed halo at $z=20$, as seen from the
$J_{21}-T_{\rm{vir}}$~plane.  $t_{\rm cool}$ is defined as the time
that the halo takes to cool from virialisation to the moment when
$T=0.75T_{\rm{vir}}$~(or, if it doesn't cool, we stop the integration
when the scale factor reaches 10).  Any halo that falls on the
\opt{colour}{red}\opt{bw}{white} region will not be able to cool in a
Hubble time.  On the other hand haloes on the dark \opt{colour}{blue
or violet}\opt{bw}{} part of the plot will cool in much less than a
dynamical time.

\begin{figure}
\begin{center}
\opt{colour}{\psfig{width=9cm,angle=90,file=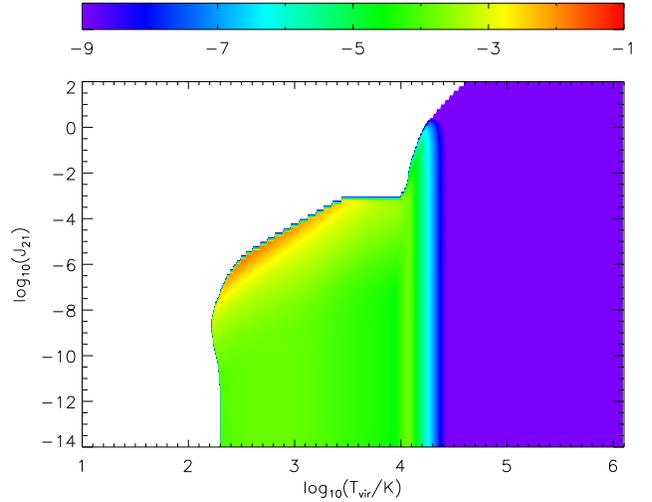}}
\opt{bw}{\psfig{width=9cm,angle=90,file=figure2_bw.ps}}
\caption{The molecular hydrogen fraction, $f_{\htwosub}$, after one
  cooling time, in the $J_{21}-T_{vir}$ plane.  The region
  corresponding to clouds that have not cooled by the time the expansion
  factor reaches 10 has been coloured white.}
\label{fig:fh2contours}
\end{center}
\end{figure}

In Figure~\ref{fig:fh2contours} we show the fractional density of
molecular hydrogen, $f_{\htwosub}$, after one cooling time in the
$J_{21}-T_{\rm{vir}}$~plane.  The region that has not cooled by the
time the scale factor reaches 10 has been coloured white.
For the zero-flux case, we see that the highest
molecular fractions occur for a virial temperature of just over
10$^4$\,K.  Halos of this temperature are able to partially ionize
hydrogen atoms and the free electrons then go on to catalyse
production of \htwo.

From these plots, we can see that the ionising radiation has two main
effects.  Firstly, it provides a heat source that prevents cooling of
halos.  The minimum virial temperature of halos that can cool
gradually increases along with the photon flux, whereas the heating
has little effect on halos above this temperature.  The effect is
visible in Figure~\ref{fig:3dcoolt} as a sharp transition between the
\opt{colour}{green/blue and red}
\opt{bw}{grey and white}
regions that runs diagonally across the plot.

Secondly, the ionizing flux can boost cooling of low-temperature halos
($\Tvir\leq10^4$\,K) by creating free electrons that then catalyse the
production of \htwo.  For any given temperature, there is only a
narrow range of photon fluxes for which this is important before
photo-heating swamps the increased \htwo\ cooling.  

Thus, as the background ionising radiation builds up in the Universe,
we expect there at first to be a small boost in the formation of
low-temperature (low-mass) star clusters and then a sharp decline.
The precise values of $J_{21}$ at which this occurs will depend upon
the redshift of structure formation (recall that the above plots are
for a redshift of 20 and the required values of $J_{21}$ will scale as
$(1+z)^3$).  In Section~\ref{sec:global} we consider both specific
values of $J_{21}=10^{-10}, 10^{-5}, 10^{-2} \&\ 10$ and also a
time-evolving ionisation field.

\section{Merger Tree}
\label{sec:tree}

In a previous paper \citep{ST03} we generated merger histories of
dark matter halos using the block model of \citet{CoK88}. In this
model, a parent block of mass M$_{0}$ and density fluctuation
$\delta_{0}$ is halved producing two daughter blocks of mass $M_{1} =
M_{0}/2$.  Extra power, drawn at random from a Gaussian distribution,
is then added to one block and subtracted from the other in order to
conserve the overall level of fluctuations in the root block.  This
process is then repeated until the desired resolution has been
reached. The reader should note that the tree is produced by stepping
back in time. Therefore, the formation of the daughter blocks occurs
at a higher redshift than their parent block. A valid criticism of
this model is that the mass distribution is not smooth, as each level
changes in mass by a factor of two.  This discretisation of mass is an
undesirable constraint on our model and for this reason we have moved
to a more realistic method of generating the merger tree.

We now use a Monte-Carlo algorithm to generate the merger tree.  There
are a number of codes which use this technique
\citep[e.g.][]{KW93,SK99} however, we have chosen the method proposed
by \citet{CLBF00}.  In short, the code uses the extended
Press-Schecter formalism to generate the merger histories of dark
matter halos, but we refer the reader to this paper for a complete
discussion of the code.  The main advantage of this method over the
block model is that the discretisation of mass has been removed.  

Unlike most previous tree implementations, we do not place our merger
tree onto a predefined grid of timesteps but utilise a very fine time
resolution. This allows us to retain the block model's prescription of
two progenitors per merger, consistent with our previous work.  The
algorithm allows us to account for the accretion of mass below the
resolution limit, however we assume that this will not seriously
affect the structure and cooling of the halo.

An important aspect of the code is how we treat the mergers of
halos.  The outcome of a merger will depend upon the ratio of masses of
the merging halos,
\begin{equation}
 q =\frac{M_1}{M_2},
\end{equation}
where $M_1 <  M_2$.  Consequently, we introduce, $q_{\rm min}$, as the
minimum mass ratio to affect the cooling of a halo.  This leads to two
cases: 
\begin{itemize}
\item If $q > q_{\rm min}$ then the two halos merge and their
 cooling is completely disrupted.  The gas is then shock heated to the
 virial temperature of the parent halo, erasing all previous cooling
 information, and the cooling starts afresh.
\item If $q < q_{\rm min}$ then we assume that the smaller of the
  two daughters is disrupted.  We then compare the times at which the
  larger daughter and the parent halo would cool.  If the former
  occurs first then we postpone the merger and allow the cooling of
  the daughter to proceed; otherwise we continue as for $q >
  q_{\rm min}$.
\end{itemize}

In the future we would like  to determine an appropriate value for
q$_{\rm min}$ from hydrodynamical  simulations.  Simulations of galaxy
mergers \citep[e.g.][]{MH96} show that when objects with mass ratios q
$>$ 0. collide the galactic structure of both objects is seriously
disrupted. They classify these as `major'  mergers.  We suspect that
smaller mass ratios  would  still  sufficiently  disrupt  the  cooling
gas  cloud. Consequently, for this work we  set q$_{\rm min}$ = 0.25.
Although we do not show it here, our results remain largely unchanged
in the range 0.2 $<$ q$_{\rm min}$ $<$ 0.3. 

We treat the metal enrichment of halos in the same way as ST03 where
it  was assumed  that,  regardless  of whether  or  not star  clusters
survive a  merger, they instantly contaminate  their surroundings with
metals and the enrichment is confined  to the next level of the merger
hierarchy  (i.e. the  metals  do  not propogate  into  halos on  other
branches).   This  is   the  same   for  both   the   global  (section
\ref{sec:global}) and local models (section \ref{sec:local}).

Once  a  halo  has  been  contaminated  it is  no  longer  classed  as
primordial, irrespective  of  whether   it  can  form  more  stars  or
not. However, no attempt is made to account for the transition from
population {\rm   III} to population {\rm II} star-formation.
Consequently, contaminated halos are assumed to cool at the same rate
as their primordial counterparts and in the case of our local model,
produce the same ionising flux. We intend to investigate the effect of
this transition in future work by including cooling from metals.  

In this paper, in common with ST03, we use a root mass of
10$^{11}$\Msun\, for our tree, and a mass resolution of
$9.5\times10^4$\,\Msun.  However, we use slightly different
cosmological parameters as derived from the WMAP data
(\citealt{SVP03}) of $\Omega_0=0.3$, $\lambda_0=0.7$,
$\Omega_{b0}=0.0457$, $h=0.71$, $\sigma_8=0.9$, and a power spectrum
as calculated by {\sc cmbfast}.

\begin{figure}
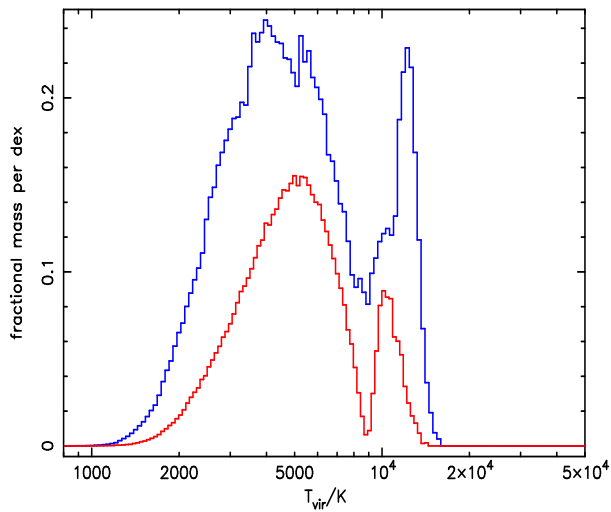

\begin{center}
\opt{colour}{\psfig{height=6.7cm,width=8cm,angle=270,file=figure3_c.ps}}
\opt{bw}{\psfig{height=6.7cm,width=8cm,angle=270,file=figure3_bw.ps}}
\caption{The fractional mass per dex of primordial star clusters as a
function of virial temperature: \opt{colour}{red}\opt{bw}{solid} line
-- new merger tree, \opt{colour}{blue}\opt{bw}{dashed} line -- Block Model.} 
\label{fig:comp}
\end{center}
\end{figure}

Figure~\ref{fig:comp} shows a comparison between the new merger tree
(\opt{colour}{red}\opt{bw}{solid} line) and the older Block Model from
ST03 (\opt{colour}{blue}\opt{bw}{dashed} line) of the
fractional mass per dex of primordial star clusters as a function of
virial temperature, averaged over a large number of realisations and
in the absence of ionising radiation. It is clear that the new tree
has had a significant affect on the number of primordial objects that
are formed. While the total number of objects that are able to cool
remains roughly constant, our new model produces only a third of the
primordial objects compared with the original, although the mass
fraction has only been reduced by half.  Qualitatively the results
remain unchanged: we still observe two generations of halos,
distinguished by their primary cooling mechanism, as discussed in
ST03.  In addition, we have removed all features associated with the
discrete mass steps (e.g.~the feature at $\sim$5000\,K in the original
model).

The smoothed mass distribution has lead to many unequal mass mergers
which were not present in the previous model thus increasing the
likelihood of contaminating large haloes with much smaller ones which
happen to cool first. Equally, the chance that haloes are involved in
mergers that disrupt their cooling is increased.  These effects
conspire to reduce the overall number of primordial objects.

\section{Global ionisation field}
\label{sec:global}

\subsection{Model}
\label{sec:global:model}

In this section, we consider the effect of a global ionisation field
that affects all halos equally.   As previously mentioned, we will
restrict ourselves to a power law ionising flux with index $\alpha=2$,
corresponding either to a quasar spectrum or that of stars of
primordial composition.  We present results for four cases of constant
normalisation: $J_{21}=10^{-10}$,  $10^{-5}$, $10^{-2}$ \&\ $10$.
These are chosen to be representative of a very low flux where the
effect on each halo is minimal; a flux which has a positive effect on
the capacity of the gas to form \htwo; and two examples of higher
amplitude fluxes that destroy \htwo.

In addition to the cases outlined above, we consider a time-dependent
build up of the background flux.   Using radiative hydrodynamical
calculations to examine the effect of the UV background on the
collapse of pre-galactic clouds, \cite{KTUSI00} modelled the evolution
of a UV background as:
\[J_{21}(z)=\left\{ \begin{array} {r@{\quad:\quad}l} e^{-\beta(z-5)}& 5\le z
 \le z_{\textrm{uv}} 
 \\ 1 & 3 \le z \le 5 \\ (\frac{1+z}{4})^{4} & 0 \le z \le
 3.  \end{array} \right.  \]
We adopt their model and fiducially fix the onset of the UV background
at $z_{\textrm{uv}} = 50$ (at which time the normalisation is
negligible).  We have added a factor of $\beta$ into this expression so
as to control the rate at which the field builds up and present
results for 3 cases: $\beta = 0.8$ (rapid), 1 (standard), and 2 (slow).  

For the latter cases especially, the global ionising flux can be
thought of as coming from pre-existing star clusters (or quasars) that
form in high-density regions of space and that are gradually ionising
the Universe around them.  For this reason, we take the mean-density
of the tree to be equal to that of the background Universe.   In
Section~\ref{sec:local}, we will consider a high-density region for
which the ionisation field is generated internally from the star
clusters that form in the tree.

\subsection{Results}
\label{sec:global:results}

In Figure~\ref{fig:global_f} we plot the fractional mass of primordial
star clusters as a function of (a) virial temperature and (b) halo
mass.  The 
\opt{colour}{colours red, blue, cyan \& green}
\opt{bw}{dash-dotted, dashed, dotted \& solid lines}
correspond to $J_{21}=10^{-10}$, $10^{-5}$, $10^{-2}$\&\ $10$,
respectively---note that the peak of the distributions do \emph{not}
move steadily from left to right as $J_{21}$ is increased.
Figure~\ref{fig:global_zsf} shows histograms of the star-formation
redshifts of the primordial clusters.

\begin{figure}
\begin{center}
\opt{colour}{\psfig{height=6.7cm,width=8cm,angle=-90,file=figure4a_c.ps}
\psfig{height=6.7cm,width=8cm,angle=-90,file=figure4b_c.ps}}
\opt{bw}{\psfig{height=6.7cm,width=8cm,angle=-90,file=figure4a_bw.ps}
\psfig{height=6.7cm,width=8cm,angle=-90,file=figure4b_bw.ps}}
\caption{Fractional mass per dex of primordial objects as a
 function of (a) virial temperature and (b) mass, for 4 different
 cases: \opt{colour}{red}\opt{bw}{dash-dotted} corresponds to
 $J_{21}=10^{-10}$; \opt{colour}{blue}\opt{bw}{dashed} to
 $J_{21}=10^{-5}$; \opt{colour}{cyan}\opt{bw}{dotted} to $J_{21}=10^{-2}$; and
 \opt{colour}{green}\opt{bw}{solid} to $J_{21}=10$.}
\label{fig:global_f}
\end{center}
\end{figure}

\begin{figure}
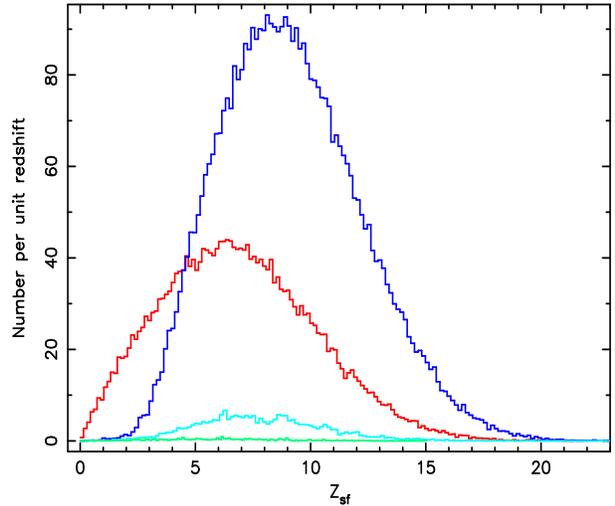

\begin{center}
\opt{colour}{\psfig{height=6.7cm,width=8cm,angle=-90,file=figure5_c.ps}}
\opt{bw}{\psfig{height=6.7cm,width=8cm,angle=-90,file=figure5_bw.ps}}
\caption{Histograms of star-formation redshifts for primordial halos.
 The \opt{colour}{colour}\opt{bw}{line-style} coding is the same as in
 Figure~\ref{fig:global_f}.}
\label{fig:global_zsf}
\end{center}
\end{figure}

The lowest amplitude case is almost indistinguishable from that of
zero flux. For this reason we have not plotted the latter. There are two
bumps in the virial temperature histogram corresponding to two
distinct cooling mechanisms.  In ST03, these were christened
Generation~1 (low-virial temperature, $T\leq8\,600$\,K, low-mass, high
collapse redshift, dominated by \htwo\ cooling) and Generation~2
(high-virial temperature, $T\geq8\,600$\,K, high-mass, low collapse
redshift, dominated by electronic cooling).  

As the flux is increased, the effect of the radiation field is to
promote the cooling of Generation~1 halos.  The typical virial
temperature and mass of such halos decreases, and the number of
Generation~2 star clusters is reduced as collapsing halos are more
likely to have been polluted by metals from smaller objects within
them.  For $J_{21}=10^{-5}$ (\opt{colour}{blue}\opt{bw}{dashed}
curve), the effect is so pronounced that it completely eliminates
Generation~2 objects.  However this is an extreme case, because, as
Figures~\ref{fig:3dcoolt} \&\ \ref{fig:fh2contours} show, this flux
has been chosen to produce close to the maximum possible enhancement
in the \htwo~fraction and a corresponding reduction in cooling time
throughout the redshifts at which these halos form.

If the background flux is increased further, then the enhancement in
Generation~1 star clusters is reversed.  For $J_{21}=10^{-2}$
(\opt{colour}{cyan}\opt{bw}{dotted} curve) the balance has shifted
almost entirely in the favour of Generation~2 clusters and by
$J_{21}=10$ all Generation~1 clusters have been eradicated.

It is interesting to note that the mass fraction of stars contained in
primordial star clusters is not greatly affected by the normalisation
of the ionising radiation, varying from 0.05 to 0.1.  However, the mass
(and hence number) of the star clusters varies substantially.  A
modest flux will increase the number of small clusters, moving the
mass function to lower masses while a greater flux produces the
opposite effect.

The number of primordial star clusters as a function of star formation
redshift is shown in figure~\ref{fig:global_zsf}. The redshift
distribution is similar in all cases, which differs significantly from
model with time-varying flux, discussed next. 

Figure \ref{fig:time_f} plots the fractional mass of primordial star
clusters, as a function of (a) virial temperature and (b) mass.
Unlike our previous results, the introduction of a time-evolving field
has had a devastating effect on the mass fraction of primordial
objects.  As the rate at which the flux builds-up increases, the peak
of the Generation~1 mass function moves towards lower masses (and
virial temperatures) and its normalisation decreases significantly.
At the same time, as shown in figure~\ref{fig:time_zsf}, the peak in
the production rate moves to higher redhsifts.  In each case, it
corresponds to a flux of $J_{21}\approx5\times10^{-5}$ for which, from
the previous results, an enhancement in the \htwo~fraction and hence a
dcrease in the cooling time is expected.

The effect on the production of higher-mass, higher-virial-temperature
star clusters, is more complicated.  In this paper we are concerned
with primordial objects, by which we mean those with zero metallicity.
The key question, then, is whether large halos are contaminated by
metals from sub-clusters within them.  With a slow build-up of flux
cooling in these sub-clusters is enhanced resulting in increased
contamination and a reduction in the number density of primoridal
Generation~2 halos.  However, a rapid build-up of flux cuts off
production of small halos dramatically and the number of primordial
Generation~2 halos is increased.

\begin{figure}
\begin{center}
\opt{colour}{
\psfig{height=7cm,width=8cm,angle=-90,file=figure6a_c.ps}
\psfig{height=7cm,width=8cm,angle=-90,file=figure6b_c.ps}}
\opt{bw}{
\psfig{height=7cm,width=8cm,angle=-90,file=figure6a_bw.ps}
\psfig{height=7cm,width=8cm,angle=-90,file=figure6b_bw.ps}}
\caption{Fractional mass per dex of primordial objects as a
 function of (a) virial temperature and (b) mass, for 4 different
 cases: \opt{colour}{orange}\opt{bw}{dash-dotted} corresponds to the
 no flux case; \opt{colour}{blue}\opt{bw}{dashed} to $\beta = 2$;
 \opt{colour}{red}\opt{bw}{dotted} to $\beta = 1$; and
 \opt{colour}{green}\opt{bw}{solid} to $\beta = 0.8$.}
\label{fig:time_f}
\end{center}
\end{figure}

\begin{figure}
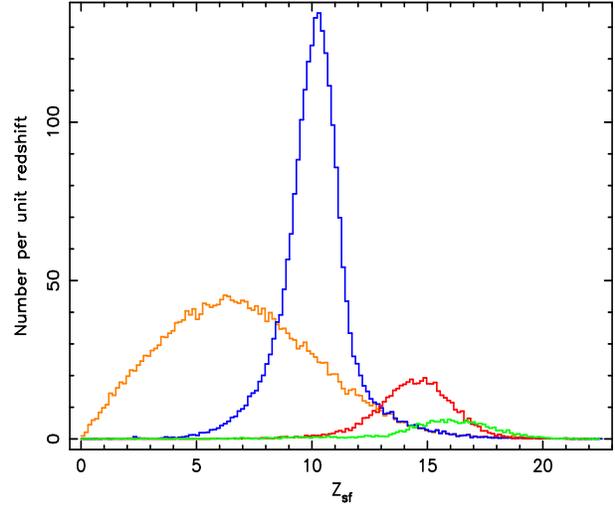

\begin{center}
\opt{colour}{\psfig{height=6.7cm,width=8cm,angle=-90,file=figure7_c.ps}}
\opt{bw}{\psfig{height=6.7cm,width=8cm,angle=-90,file=figure7_bw.ps}}
\caption{Histograms of star-formation redshifts for primordial halos.
 The \opt{colour}{colour}\opt{bw}{line-style} coding is the same as in
 Figure~\ref{fig:time_f}.}
\label{fig:time_zsf}
\end{center}
\end{figure}

\section{Local feedback}
\label{sec:local}

In this section, we consider local feedback, i.e.~that produced by
star clusters internal to the tree.

\subsection{Model}
\label{sec:local:model}

\subsubsection{Physical picture and assumptions}
\label{sec:local:model:picture}

The large value of the Thomson electron-scattering optical depth in
the WMAP data of $\tau=0.17\pm0.04$ \citep{SVP03} suggests an early
reionisation era at $11<z<30$~($180^{+220}_{-80}$Myr after the Big
Bang).  This requires a high efficiency of ionising photon production
in the first stars, corresponding to a deficit of low-mass stars.
This is backed up by theoretical arguments and simulations
\citep[e.g.~][]{ABN02,BCL02,Sch02,TVS04,SS05} of star-formation in a
primordial gas, for which fragmentation was found to be strongly
inhibited by inefficient cooling at metallicities below about
$10^{-3.5}$\,\Zsun. 

There is still some debate as to whether the first star forming halos
will produce a single massive star \citep[e.g.~][]{ABN02} or fragment
further to form the first star clusters
\citep[e.g.~][]{BCL99}. Whichever of these is correct makes little
difference to our results. \cite{TVS04} considered a number of IMFs
that may lead to the required early reionisation. These have ionising
fluxes per unit mass in the range
$Q\approx10^{47}$--$10^{48}$\,ph\,s$^{-1}\,\Msun^{-1}$, but they all
have similar ionising efficiencies of about 80\,000\,photons/baryon
when integrated over the whole life of the stars.A value which is
similar to that obtained for single massive stars ($M \gta 20
\Msun$). If we assume a uniform flux over time, then this corresponds
to mean lifetimes of $3.0\times10^7$--$3.0\times10^6$\,yr,
respectively. Note that these values of $Q$ are much greater than the
average for the Milky Way,
$Q\approx8.75\times10^{43}$\,ph\,s$^{-1}\,\Msun^{-1}$ \citep{RS00}.  

When halos in the merger tree are able to cool, we assume that they
will convert part of their baryonic component into a ``star cluster''
(primordial or otherwise). These objects will exert radiative feedback
onto the next generation of haloes that form inside the same tree. The
photon flux will also depend upon the star-formation efficiency and
the escape fraction from the star-forming region in the centre of the
halo.  In this paper, we are not concerned with the magnitude of metal
production and so it is only the combination of the two, $f_*$, that
is of interest. 

The total ionising flux from a halo is
\begin{equation}
S=f_*\,M_b\,Q,
\end{equation}
where $M_b\approx0.152\,M$ is the baryonic mass and $M$ is the total
mass of the halo.  We will present results for 3 models, listed in
table~\ref{tab:local:model}.  Model~1 has the highest ionising flux;
model 2 has a smaller flux but lasts for the same length of time;
model 3 has an even smaller flux but lasts for longer so that the
total number of ionising photons produced is the same as for model 2.

\begin{table}
\begin{center}
\caption{Parameters of the ionisation models that we consider: model
number; fraction of mass in stars, $f_*$; specific ionising fluxes,
$Q_0$, in units of ph\,s$^{-1}\,\Msun^{-1}$; total number of ionising
photons, $N_0$, and ionising flux, $S_0$, in units of ph\,s$^{-1}$ per
solar mass of matter (baryonic plus dark); and the time for which the
ionising flux acts, $t_*$, in units of years.}
\label{tab:local:model}
\begin{tabular}{lccccc}
\bf{id} & $f_*$ & $Q_0$ & $N_0$ & $S_0$ & $t_*$ \\
\hline
1 (\opt{colour}{red}\opt{bw}{solid})& $10^{-2}$ & $10^{48}$ & 
$1.5\times10^{59}$ & $1.5\times10^{45}$ & $3\times10^6$ \\
2 (\opt{colour}{green}\opt{bw}{dashed})& $10^{-3}$ & $10^{48}$ & 
$1.5\times10^{58}$ & $1.5\times10^{44}$ & $3\times10^6$ \\
3 (\opt{colour}{blue}\opt{bw}{dotted})& $10^{-3}$ & $10^{47}$ & 
$1.5\times10^{58}$ & $1.5\times10^{43}$ & $3\times10^7$
\end{tabular}
\end{center}
\end{table}

Once the stars have formed, the ionising photons will begin to
evaporate the rest of the halo and make their way into the surrounding
IGM.  Each star cluster produces
\begin{equation}
N_\gamma=80\,000\,{f_*\,M_b\over m_{\rm H}}
\end{equation}
ionising photons, where $M_b$ is the baryonic mass.
In the absence of recombination, this is sufficient to to ionise the
hydrogen in a region of baryonic mass
\begin{equation}
M_{b\gamma}=N_\gamma\,\mu_{\rm H}m_{\rm H},
\end{equation}
where $\mu_{\rm H}\approx1.36$ is the mean molecular mass per hydrogen
nucleus.  For the star-formation efficiencies and top-heavy
initial-mass function that we consider here, there are more than
enough photons to ionise any neutral gas within the star-cluster:
\begin{equation}
{M_{b\gamma}\over M_b}=80\,000\,f_*\,\mu_{\rm H}\approx
1.1\times10^5\,f_*.
\end{equation}

We next consider whether it is correct to neglect recombinations.
The photon flux required to maintain ionisation of the halo (at the
mean halo density) is given by 
\begin{equation}
S_{\rm halo}={4\pi\over3}\,R^3\,n_{\rm{H}}^2\,\mathcal{R},
\label{eq:absorb}
\end{equation}
where $R$ is the radius of the virialised halo, $n_{\rm{H}}$ is the
combined number density of all species of hydrogen, and
$\mathcal{R}$ is the recombination rate \citep[see,
e.g.~][]{HST02}.  The value of $S_{\rm halo}$ would be higher if we
were to take into account clumping of the gas.  On the other hand,
there are two effects that will tend to lower $S_{\rm halo}$: for
high-temperature halos not all the gas will be neutral; for
low-temperature halos the gas will be raised to a temperature that
exceeds the virial temperature and so will tend to escape from the
halo---the sound-crossing time for a gas at 10$^4$K is of order
$1.0\times10^7$yr for a $10^6$\Msun~halo at an expansion factor
$a=0.05$.  We assume that these effects will roughly cancel and
set the nett photon flux that escapes the halo equal to $S_{\rm
  esc}=S-S_{\rm halo}=f_{\rm esc}\,S$.  Here
\begin{equation}
1-f_{\rm esc}={S_{\rm halo}\over
  S}\approx0.12\,\left(a\over0.05\right)^{-3}\, 
\left(S_0\over 1.5\times10^{44}\mbox{ph\,s}^{-1}\right)^{-1},
\label{eq:escape}
\end{equation}
where we have set $\mathcal{R}$ equal to the recombination rate for a
$10^4$K gas.  Of course, $f_{\rm esc}$ is not allowed to drop below zero.
The number of ionising photons that escape the source halo is
\begin{equation}
N_{b\gamma,{\rm esc}}=f_{\rm esc}N_{b\gamma}.
\end{equation}

Escaping photons are now free to propagate into the inter-halo medium
(aka inter-galactic medium or IGM) and irradiate nearby halos.  At the
mean density of the IGM, the Str\"omgren radii for any ionised regions
are very large, and only model~1 produces enough photons to ionise out
to the Str\"omgren radius, and then only at very early times,
$a\approxlt0.02$. 
A better picture is that of a bubble of ionised gas whose outer radius
grows with time until the ionising source switches off.  To a good
approximation then, and for simplicity, we assume that recombinations
in the IGM are negligible.

\subsubsection{Numerical methodology}
\label{sec:numericalm}

\begin{figure}
\begin{center}
\psfig{width=8cm,angle=0,file=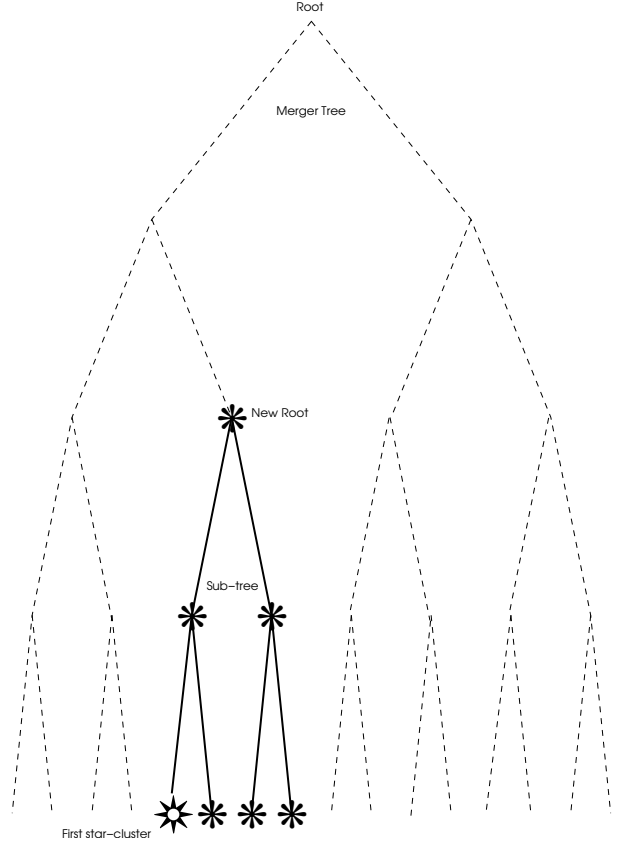}
\caption{Schematic view of the merger tree under internal
 feedback.  Once the first star cluster is located, the code calculates
 how many levels up the tree has to go to re-calculate the cooling
 times of each halo under the new root, this time under the influence
 of the radiative flux coming from that first star cluster.  In this
 example, the feedback region reaches two levels up the tree.  Then,
 the whole sub-tree cooling times will be re-evaluated.}
\label{fig:subtree}
\end{center}
\end{figure}

The local feedback is implemented as follows.  First the tree is
scanned for all star clusters and a list is generated, in order of
decreasing star formation redshift.  Starting with the first cluster,
we work up the tree looking at the baryonic mass of successive parent
halos, $M_{b,{\rm par}}$ until the last halo for which
\begin{equation}
M_{b,{\rm par}}<M_{b}+M_{b\gamma,{\rm esc}}.
\label{eq:par_mass}
\end{equation}
The sub-tree below this parent halo (see Figure~\ref{fig:subtree})
defines the extent of the ionised region and the cooling times of all
halos within it are recalculated taking into account the amplitude of
the flux and the time for which it acts.

The tree provides limited information about the spatial distribution
of halos.  However, we know that halos are confined within a common
parent halo and that the parent will not have collapsed at this time.
So we take the separation of the star cluster and each neighbouring halo
to be equal to the radius of a sphere at the mean density at that time
that encloses a mass equal to that of the first common parent:
\begin{equation}
R_{\rm{par}}=a\,\left(3~M_{b,{\rm par}}\over4\,\pi\,\Omega_{b0}
  \,\rho_{c0}\right)^{1/3}
\label{eq:par_radius}
\end{equation}
where $a$~is the scale factor at the time of star formation and the
other quantities have the usual meanings.  Note that the value of
$M_{b,{\rm par}}$ in equation~\ref{eq:par_radius} will vary depending
upon how far one has to travel up the tree to find a common neighbour.


The flux density, $F$, at a distance $R_{\rm par}$~from the source is
given by:
\begin{equation}
F={S_{\rm esc}\over{4\pi R_{\rm{par}}^2}}
\label{eq:jdist}
\end{equation}
We need to convert this into an equivalent value of $J_{21}$ for input
into our chemical evolution code.  To do this we integrate the
spectrum given in equation~\ref{eq:spectrum} over all frequencies and
angles
\begin{equation}
F=4\pi\int_{\nu_H}^\infty{J_\nu\over{h\nu}}d\nu
={4\pi\over\alpha}\,J_{21},
\label{eq:jspec}
\end{equation}
where $h\nu_H=13.6$eV is the energy of \h~ionisation.
Combining these two equations allows us to express the ionising flux
as an equivalent value of $J_{21}$ for an isotropic radiation field
(the directionality of the radiation field is unimportant in this context).
We use the value of $J_{21}$ and duration of the ionisation as inputs
to the chemistry code to obtain the new cooling time of each
neighbouring halo in the sub-tree, using the heating and cooling
processes explained in Section \ref{sec:isocool}.

The list of remaining clusters is then re-ordered using the new
star-formation redshifts and we continue with the next halo in the
list.  The process is repeated until we reach the bottom of the list or
until the next halo in the list is not able to cool in a Hubble time.

\subsection{Results}
\label{sec:local_result}

The first stars are widely predicted to form in high density regions
of space.  Consequently, in this section we present results of
simulations with internal radiative feedback in a region for which the
root halo corresponds to a positive 3-$\sigma$ density fluctuation.

\begin{figure}
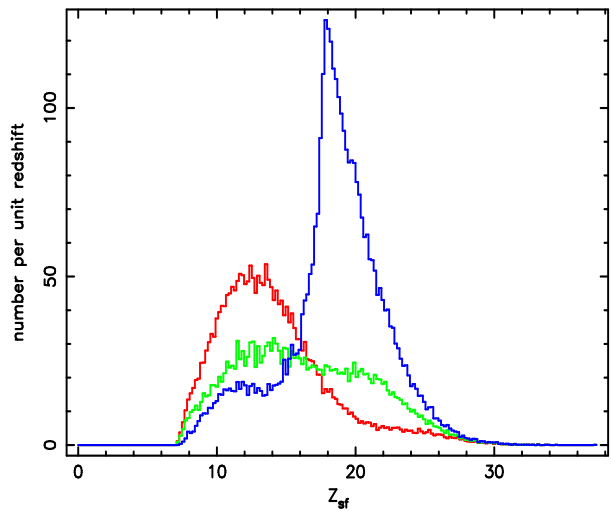

\begin{center}
\opt{colour}{\psfig{height=6.7cm,width=8.cm,angle=-90,file=figure9_c.ps}}
\opt{bw}{\psfig{height=6.7cm,width=8.cm,angle=-90,file=figure9_bw.ps}}
\caption{Histograms of the star formation redshifts of primordial star
 clusters averaged over a large number of realisations.  The
 \opt{colour}{colours}\opt{bw}{line styles} correspond to different
 models of local feedback as listed in table \ref{tab:local:model}.}
\label{fig:local_zsf}
\end{center}
\end{figure}

Figure \ref{fig:local_zsf} plots histograms of star formation
redshifts for primordial halos for the three  cases shown in Table
\ref{tab:local:model}, averaged over a large number of
realisations. It is evident that the redshift evolution is markedly
different for the three curves and we shall discuss each in turn. 

\begin{figure}
\begin{center}
\opt{colour}{\psfig{height=6.7cm,width=8.cm,angle=-90,file=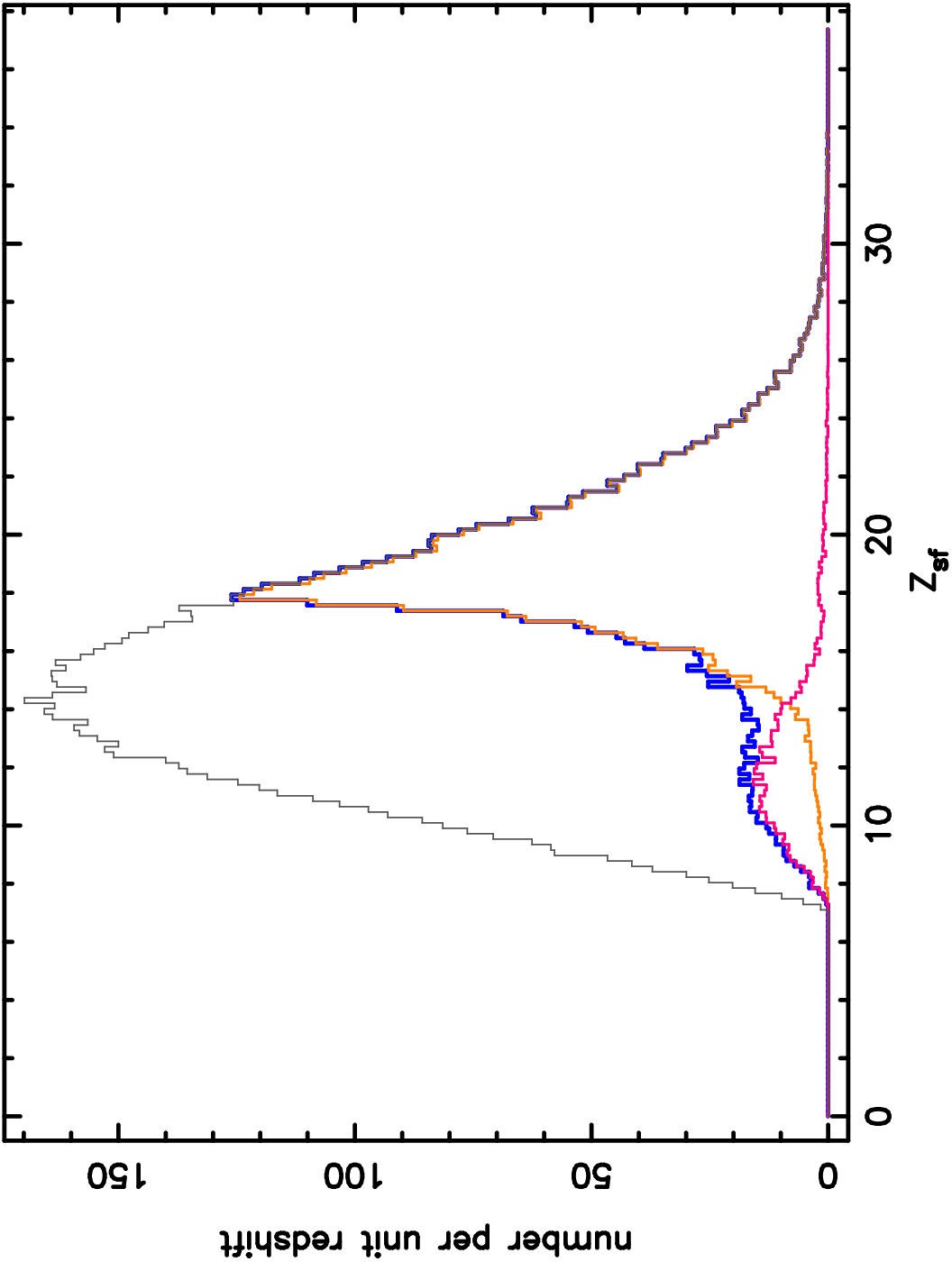}}
\opt{bw}{\psfig{height=6.7cm,width=8.cm,angle=-90,file=figure10_bw.ps}}
\caption{Histogram of star-formation redshifts for primordial
 haloes. Here we compare model 3 (\opt{colour}{blue}\opt{bw}{thick
 solid line}) with the the no flux case
 (\opt{colour}{grey}\opt{bw}{thin solid line}).  Also show are the
 contributions from Generation 1
 (\opt{colour}{orange}\opt{bw}{dashed}) and Generation 2
 (\opt{colour}{magenta}\opt{bw}{dotted}) halos.}
\label{fig:local_gen}
\end{center}
\end{figure}

First of all, in figure \ref{fig:local_gen}, we compare model 3 with
the case of zero flux. We can clearly see that the evolution of the
two curves is identical up to a point, after which model~3 drops away
dramatically.  This sudden change can be understood by examining
equation \ref{eq:escape}. We are interested in the redshift at which
the ionising photons first begin to escape the halo.  If we set
$f_{\rm esc}$ equal to zero and solve for the scale factor, a, we find
that, for this particular model, photons do not break out until a
redshift $z \sim 18$, in agreement with what is seen in figure
\ref{fig:local_gen}. Also shown are the contributions from the two
different halo generations, discussed in previous sections.  The
escaping photons have had a devastating impact on the surrounding
halos, particularly the smaller, Generation 1 halos. As a result there
is a rise in the number of Generation 2 halos because the reduced
contamination at early times allows more massive halos to cool as
primordial objects.

In models 1 \& 2 the photons are able to escape the halo at much
earlier times (before any objects have been able to cool). As such,
the first primordial objects to form will immediately begin to
influence their surroundings. This explains the much greater reduction
in Generation~1 star clusters for these models compared to model 3,
seen in figure \ref{fig:local_zsf}.

\begin{figure}
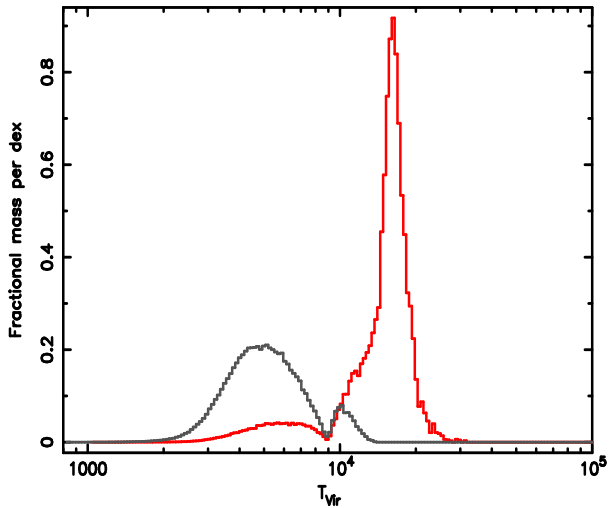

\begin{center}
\opt{colour}{\psfig{height=6.7cm,width=8.cm,angle=-90,file=figure11_c.ps}}
\opt{bw}{\psfig{height=6.7cm,width=8.cm,angle=-90,file=figure11_bw.ps}}
\caption{Fractional mass per dex of primordial objects as a function
 of temperature, for model 1 (\opt{colour}{red}\opt{bw}{thick line})
 and the no flux case (\opt{colour}{grey}\opt{bw}{thin line}).}
\label{fig:local_tempf}
\end{center}
\end{figure}

Figure \ref{fig:local_tempf} plots the fractional mass per dex of
primordial objects as a function of temperature for our highest flux
case (model 1). As expected, the higher flux has suppressed the
the small, low temperature halos at high redshift, thus reducing the
amount of early contamination. Once again we see that there has been
an enhancement in the number of high temperature, high mass halos, the
distribution of which is reminiscent of that seen in figure
\ref{fig:global_f} for the high values of $J_{\rm 21}$ in our global
model, particularly $J_{\rm 21} = 10^{-2}$.  Indeed, the average
$J_{\rm 21}$ values received by halos in this model are in the range
$J_{\rm 21} = 10^{-3}-10^{-2}$, consistent with our previous results.

With ten times fewer photons, we expect model~2 to be less destructive
than model~1 at higher redshifts and figure \ref{fig:local_zsf}
confirms this fact. The \opt{colour}{green}\opt{bw}{dashed} curve
shows more primordial objects early on which consequently reduces the
number of Generation 2 objects that form. Interestingly, this model
shows an approximate balance between the two generations with a
roughly constant formation rate of primordial halos between
redshifts $z \sim 10-22$.

The mass fraction of stars contained in primordial star
clusters for all the models presented here remains relatively
constant, varying from 0.06 to 0.13.  However, the mass functions vary
substantially as do the star formation histories.  As the specific
ionizing flux increases, the balance moves towards later star
formation and more Generation~2 primordial star clusters.  The
positive feedback into Generation~1 clusters seen in
figure~\ref{fig:global_f} for intermediate values of $J_{21}$ lasts
for too short a time to be noticable.

\section{Discussion}
\label{sec:raddiscuss}

This paper looks at the impact of radiative feedback on primordial
structure formation. This is done in two ways.

The first part investigates the properties of primordial objects under
a global UV background.  The merger tree is illuminated by a constant
and isotropic radiation field of four different intensities,
parametrised by a constant value of $J_{21}$: $10^{-10}$, $10^{-5}$,
$10^{-2}$ and $10$.  It seems more plausible that any background
radiation field would gradually build up over time with the formation
of more and more primordial stars.  Consequently, we also investigate
a time-dependent build-up of the background flux using an extension of
the model of \citet{KTUSI00}.  The section of the paper uses a
mean-density region of space as the background radiation field is
assumed to come from external sources within higher-density regions.

The effect of a constant UV field on the halo population is not a
trivial one as both positive or negative feedback can arise from
different choices of the flux amplitude $J_{21}$. The cooling of a
low-temperature primordial gas is almost completely dominated by the
release of energy from roto-vibrational line excitation of \htwo. But
if a radiation field is present, \htwo~is easily destroyed by
Lyman-Werner photons ($11.2$eV$<h\nu<13.6$eV).  On the other hand, the
formation of \htwo~can be enhanced by an increase in the ionization
fraction produced by a weak ionising flux, as electrons act as a
catalyst for the formation path of \htwo.  At modest flux levels of
$J_{21}\sim10^{-5}$ the nett effect is to boost cooling in the first
star clusters.  A similar result has previously been found by
\cite{HRL96}, \cite{RGS01}, \cite{KSUI01}~and \cite{YAHS03}.

Negative feedback is produced not only by photo-dissociation of \htwo;
at high flux levels the dominant effect comes from heating associated
with photo-ionisation of \h.  For fluxes of $J_{21}\geq10^{-2}$ we
find that molecular cooling is ineffective and only haloes with virial
temperatures of $T_{\rm{vir}}>14000$K are able to cool in a Hubble
time.  Because of reduced contamination from star-formation in
low-mass halos, strong radiation fields can increase the number of
high-mass primordial star clusters.

The second part of this paper dealt with a model in which the
radiative feedback is localised. That is, star clusters irradiate
their surrounding area, changing the cooling properties of those
primordial objects that are inside their ionisation spheres. For this
model we considered only a high density region corresponding to a
3-$\sigma$ fluctuation since the first objects are thought to form in
regions of high overdensity.

The effect of the feedback depends mainly upon the specific ionising
flux, averaged over the mass of the halo.  When this is low, most or
all of the photons will be used up in maintaining the ionisation of
the halo.  For the particular model that we describe in this paper,
equation~\ref{eq:escape} relates the escape fraction, $f_{\rm esc}$, of
ionising photons to the specific ionising flux, $S_0$, and redshift.
A value of $S_0$ below $10^{43}$ph\,s$^{-1}$\,\Msun$^{-1}$ will reduce
$f_{\rm esc}$ to zero until the redshift drops to about 16,
corresponding to the peak in the production rate of primordial halos
per unit redshift.  Although the precise number will be model
dependent, we regard this as a fiducial value below which feedback
will be ineffective.

Higher values of $S_0$ result in shift from primordial star clusters
away from Generation~1 (low virial temperature) towards Generation~2
(high virial temperature).  Unlike the case of a global ionisation
field, Generation~1 clusters are not erradicated completely, because
some must form in order to provide the feedback.  However, a specific
ionisation flux of $S_0=10^{45}$ ph\,s$^{-1}$\,\Msun$^{-1}$ is enough
to swing the balance strongly in favour of Generation~2 star clusters.

This paper makes a number of advances on our previous modelling, most
notably the use of an improved merger tree that does not restrict
halos to factors of two in mass, and the introduction of a radiation
field.  The former results in a reduced mass fraction of stars in
primordial halos with the bias shifting more in favour of
Generation~1; however, the latter moves the bias back the other way.
The conclusions of HSTC02 and ST03 remain valid in that there could be
a substantial population of primordial star clusters that form in
high-mass halos dominated via electronic cooling.

Further improvements to our model are possible.  Although we do not
expect these to change our qualitative conclusions, they will be
important for making precise quantitative predictions about the number
density and composition of the first star clusters.  We mention some
of them below.

We assume in this paper that the internal structure of halos is
unaltered between major merger events.  However, it is possible for
halos to increase their mass substantially through a succession of
minor accretion events, and this will release gravitational potential
energy and lead to heating \citep{YAHS03, RBF05}.  We intend to
incorporate this in future work.

No attempt has been made to distinguish between star-formation and
feedback in primordial star clusters and later generations.  It is
expected that the initial mass function should alter considerably once
the metallicity reaches about $10^{-3.5}$\,Z$_\odot$ \citep{BrL03,
  SFN02,SS05} and that the spectrum of the ionization field would become
softer and its normalisation lower.

We have neglected the effect of stellar winds and supernovae, both of
which help to heat the surrounding gas and pollute it with metals.
The effect of winds is fairly localised, but supernovae can lead to
superwinds that expel enriched material from the star cluster.  It is
possible that this could affect neighbouring halos; more likely it
will simply ensure thorough mixing of metals throughout a common
parent halo.  Numerical simulations will be required in order to model
this process with any degree of realism.

Finally, it would be valuable as computing resources improve to
simulate larger regions of space that would provide a more
respresentative section of the Universe.  This would allow us to
distinguish between the early history of galaxies that are located
within clusters and in mean density regions of space, for example.  It
would also allow for both localised feedback and a self-consistent
build-up of the background, global radiation field.

\section*{Acknowledgements}
MAM is supported by a PPARC studentship.

\bibliographystyle{mn2e} 
\bibliography{early}

\begin{thebibliography}{}

\bibitem[\protect\citeauthoryear{Abel, Anninos, Norman \& Zhang}{Abel
  et~al.}{1998}]{AAN98}
Abel T.,  Anninos P.,  Norman M.~L.,    Zhang Y.,  1998, ApJ, 508, 518

\bibitem[\protect\citeauthoryear{{Abel}, {Anninos}, {Zhang} \& {Norman}}{{Abel}
  et~al.}{1997}]{AAZN97}
{Abel} T.,  {Anninos} P.,  {Zhang} Y.,    {Norman} M.~L.,  1997, New Astronomy,
  2, 181

\bibitem[\protect\citeauthoryear{{Abel}, {Bryan} \& {Norman}}{{Abel}
  et~al.}{2002}]{ABN02}
{Abel} T.,  {Bryan} G.~L.,    {Norman} M.~L.,  2002, Science, 295, 93

\bibitem[\protect\citeauthoryear{{Barkana} \& {Loeb}}{{Barkana} \&
  {Loeb}}{1999}]{BL99}
{Barkana} R.,  {Loeb} A.,  1999, ApJ, 523, 54

\bibitem[\protect\citeauthoryear{Bromm, Coppi \& Larson}{Bromm
  et~al.}{1999}]{BCL99}
Bromm V.,  Coppi P.~S.,    Larson R.~B.,  1999, ApJl, 527, L5

\bibitem[\protect\citeauthoryear{Bromm, Coppi \& Larson}{Bromm
  et~al.}{2002}]{BCL02}
Bromm V.,  Coppi P.~S.,    Larson R.~B.,  2002, ApJ, 564, 23

\bibitem[\protect\citeauthoryear{Bromm \& Loeb}{Bromm \& Loeb}{2003}]{BrL03}
Bromm V.,  Loeb A.,  2003, Nat., 425, 812

\bibitem[\protect\citeauthoryear{{Cen}}{{Cen}}{2003}]{Cen03}
{Cen} R.,  2003, ApJ, 591, L5

\bibitem[\protect\citeauthoryear{{Ciardi}, {Ferrara} \& {White}}{{Ciardi}
  et~al.}{2003}]{CFW03}
{Ciardi} B.,  {Ferrara} A.,    {White} S.,  2003, MNRAS, 344, L7

\bibitem[\protect\citeauthoryear{Cole \& Kaiser}{Cole \& Kaiser}{1988}]{CoK88}
Cole S.,  Kaiser N.,  1988, MNRAS, 233, 637

\bibitem[\protect\citeauthoryear{{Cole}, {Lacey}, {Baugh} \& {Frenk}}{{Cole}
  et~al.}{2000}]{CLBF00}
{Cole} S.,  {Lacey} C.~G.,  {Baugh} C.~M.,    {Frenk} C.~S.,  2000, MNRAS, 319,
  168

\bibitem[\protect\citeauthoryear{{de Jong}}{{de Jong}}{1972}]{D72}
{de Jong} T.,  1972, A\&A, 20, 263

\bibitem[\protect\citeauthoryear{{Glover} \& {Brand}}{{Glover} \&
  {Brand}}{2001}]{GB01}
{Glover} S.~C.~O.,  {Brand} P.~W.~J.~L.,  2001, MNRAS, 321, 385

\bibitem[\protect\citeauthoryear{{Haiman}, {Abel} \& {Rees}}{{Haiman}
  et~al.}{2000}]{HAR00}
{Haiman} Z.,  {Abel} T.,    {Rees} M.~J.,  2000, ApJ, 534, 11

\bibitem[\protect\citeauthoryear{{Haiman}, {Rees} \& {Loeb}}{{Haiman}
  et~al.}{1996}]{HRL96}
{Haiman} Z.,  {Rees} M.~J.,    {Loeb} A.,  1996, ApJ, 467, 522

\bibitem[\protect\citeauthoryear{Hutchings, Santoro, Thomas \&
  Couchman}{Hutchings et~al.}{2002}]{HST02}
Hutchings R.~M.,  Santoro F.,  Thomas P.~A.,    Couchman H. M.~P.,  2002,
  MNRAS, 330, 927

\bibitem[\protect\citeauthoryear{{Kauffmann} \& {White}}{{Kauffmann} \&
  {White}}{1993}]{KW93}
{Kauffmann} G.,  {White} S.~D.~M.,  1993, MNRAS, 261, 921

\bibitem[\protect\citeauthoryear{{Kitayama}, {Susa}, {Umemura} \&
  {Ikeuchi}}{{Kitayama} et~al.}{2001}]{KSUI01}
{Kitayama} T.,  {Susa} H.,  {Umemura} M.,    {Ikeuchi} S.,  2001, MNRAS, 326,
  1353

\bibitem[\protect\citeauthoryear{{Kitayama}, {Tajiri}, {Umemura}, {Susa} \&
  {Ikeuchi}}{{Kitayama} et~al.}{2000}]{KTUSI00}
{Kitayama} T.,  {Tajiri} Y.,  {Umemura} M.,  {Susa} H.,    {Ikeuchi} S.,  2000,
  MNRAS, 315, L1

\bibitem[\protect\citeauthoryear{{Machacek}, {Bryan} \& {Abel}}{{Machacek}
  et~al.}{2001}]{MBA01}
{Machacek} M.~E.,  {Bryan} G.~L.,    {Abel} T.,  2001, ApJ, 548, 509

\bibitem[\protect\citeauthoryear{{Mihos} \& {Hernquist}}{{Mihos} \&
  {Hernquist}}{1996}]{MH96}
{Mihos} J.~C.,  {Hernquist} L.,  1996, ApJ, 464, 641

\bibitem[\protect\citeauthoryear{{Oh} \& {Haiman}}{{Oh} \&
  {Haiman}}{2002}]{OH02}
{Oh} S.~P.,  {Haiman} Z.,  2002, ApJ, 569, 558

\bibitem[\protect\citeauthoryear{{Omukai}}{{Omukai}}{2001}]{Omu01}
{Omukai} K.,  2001, ApJ, 546, 635

\bibitem[\protect\citeauthoryear{{O'Neil} \& {Reinhardt}}{{O'Neil} \&
  {Reinhardt}}{1978}]{OR78}
{O'Neil} S.~V.,  {Reinhardt} W.,  1978, J. Chem. Phys., 69, 2126

\bibitem[\protect\citeauthoryear{Osterbrock}{Osterbrock}{1989}]{O89}
Osterbrock D.~E.,  1989, Astrophysics of gaseous nebulae and active galactic
  nuclei.
Mill Valley:University Science Books

\bibitem[\protect\citeauthoryear{Reed, Bower, Frenk, Gao, Jenkins, Theuns \&
  White}{Reed et~al.}{2005}]{RBF05}
Reed D.,  Bower R.,  Frenk C.~S.,  Gao L.,  Jenkins A.,  Theuns T.,    White S.
  D.~M.,  2005, The First Generation of Star-Forming Haloes, MNRAS, submitted
  (astro-ph/0504038)

\bibitem[\protect\citeauthoryear{{Ricotti}, {Gnedin} \& {Shull}}{{Ricotti}
  et~al.}{2001}]{RGS01}
{Ricotti} M.,  {Gnedin} N.~Y.,    {Shull} J.~M.,  2001, ApJ, 560, 580

\bibitem[\protect\citeauthoryear{{Ricotti}, {Gnedin} \& {Shull}}{{Ricotti}
  et~al.}{2002}]{RGS02}
{Ricotti} M.,  {Gnedin} N.~Y.,    {Shull} J.~M.,  2002, ApJ, 575, 33

\bibitem[\protect\citeauthoryear{{Ricotti} \& {Shull}}{{Ricotti} \&
  {Shull}}{2000}]{RS00}
{Ricotti} M.,  {Shull} J.~M.,  2000, ApJ, 542, 548

\bibitem[\protect\citeauthoryear{{Santoro} \& {Shull}}{{Santoro} \&
  {Shull}}{2005}]{SS05}
{Santoro} F.,  {Shull} J.~M.,  2005, Critical metallicity and fine structure
  emission of primordial gas enriched by the first stars, ApJ, submitted
  (astro-ph/0509101)

\bibitem[\protect\citeauthoryear{{Santoro} \& {Thomas}}{{Santoro} \&
  {Thomas}}{2003}]{ST03}
{Santoro} F.,  {Thomas} P.~A.,  2003, MNRAS, 340, 1240

\bibitem[\protect\citeauthoryear{{Schaerer}}{{Schaerer}}{2002}]{Sch02}
{Schaerer} D.,  2002, A\&A, 382, 28

\bibitem[\protect\citeauthoryear{{Schneider}, {Ferrara}, {Natarajan} \&
  {Omukai}}{{Schneider} et~al.}{2002}]{SFN02}
{Schneider} R.,  {Ferrara} A.,  {Natarajan} P.,    {Omukai} K.,  2002, ApJ,
  571, 30

\bibitem[\protect\citeauthoryear{{Somerville} \& {Kolatt}}{{Somerville} \&
  {Kolatt}}{1999}]{SK99}
{Somerville} R.~S.,  {Kolatt} T.~S.,  1999, MNRAS, 305, 1

\bibitem[\protect\citeauthoryear{{Spergel}, {Verde}, {Peiris}, {Komatsu},
  {Nolta}, {Bennett}, {Halpern}, {Hinshaw}, {Jarosik}, {Kogut}, {Limon},
  {Meyer}, {Page}, {Tucker}, {Weiland}, {Wollack} \& {Wright}}{{Spergel}
  et~al.}{2003}]{SVP03}
{Spergel} D.~N.,  {Verde} L.,  {Peiris} H.~V.,  {Komatsu} E.,  {Nolta} M.~R.,
  {Bennett} C.~L.,  {Halpern} M.,  {Hinshaw} G.,  {Jarosik} N.,  {Kogut} A.,
  {Limon} M.,  {Meyer} S.~S.,  {Page} L.,  {Tucker} G.~S.,  {Weiland} J.~L.,
  {Wollack} E.,    {Wright} E.~L.,  2003, ApJ Supp., 148, 175

\bibitem[\protect\citeauthoryear{Tegmark, Silk, Rees, Blanchard, Abel \&
  Palla}{Tegmark et~al.}{1997}]{TSR97}
Tegmark M.,  Silk J.,  Rees M.~J.,  Blanchard A.,  Abel T.,    Palla F.,  1997,
  ApJ, 474, 1

\bibitem[\protect\citeauthoryear{{Tumlinson} \& {Shull}}{{Tumlinson} \&
  {Shull}}{2000}]{TS00}
{Tumlinson} J.,  {Shull} J.~M.,  2000, ApJ, 528, L65

\bibitem[\protect\citeauthoryear{Tumlinson, Shull \& Venkatesan}{Tumlinson
  et~al.}{2003}]{TSV03}
Tumlinson J.,  Shull J.~M.,    Venkatesan A.,  2003, ApJ, 584, 608

\bibitem[\protect\citeauthoryear{{Tumlinson}, {Venkatesan} \&
  {Shull}}{{Tumlinson} et~al.}{2004}]{TVS04}
{Tumlinson} J.,  {Venkatesan} A.,    {Shull} J.~M.,  2004, ApJ, 612, 602

\bibitem[\protect\citeauthoryear{{Yoshida}, {Abel}, {Hernquist} \&
  {Sugiyama}}{{Yoshida} et~al.}{2003}]{YAHS03}
{Yoshida} N.,  {Abel} T.,  {Hernquist} L.,    {Sugiyama} N.,  2003, ApJ, 592,
  645

\bibitem[\protect\citeauthoryear{{Zheng}, {Kriss}, {Telfer}, {Grimes} \&
  {Davidsen}}{{Zheng} et~al.}{1997}]{ZKT97}
{Zheng} W.,  {Kriss} G.~A.,  {Telfer} R.~C.,  {Grimes} J.~P.,    {Davidsen}
  A.~F.,  1997, ApJ, 475, 469

\end{thebibliography}

\appendix
\label{sec:appendix}

The appendix below describes how to go from cross-sections to reaction
rates and heating terms.

\section{Photo-ionisation and photo-dissociation integrals}
\label{photointegrals}
The rate at which photo-ionisation or photo-dissociation
reactions occur is given by:
\begin{equation}
k_i=4\pi\int_{\nu_{\rm th}}^\infty \sigma_{\nu,
i}\,{J_\nu\over{h\nu}}\,{\rm d}\nu,
\label{eq:rates}
\end{equation}
where $h\nu_{\rm th}$ is the threshold energy for which
photo-ionisation (or photo-dissociation) is possible and $\sigma_{\nu,
i}$~is the frequency dependent cross-section of the $i^{\rm
th}$~reaction. The cross-section and the threshold energies were taken
from Table \ref{radtab}. $J_\nu$~is the specific intensity given in
Equation \ref{eq:spectrum}.

\section{Heating terms}
\label{hterms}
The energy per particle that a photon of energy $h\nu$~transfered to
an electron in an atom or ion is $h\nu-h\nu_{\rm th}$. Therefore, the
energy per second per unit volume transferred to the gas (heating
terms or heating functions) is:
\begin{equation}
\Lambda_{{\rm heat},i}=4\pi n_i\int_{\nu_{\rm th}}^\infty
\sigma_{\nu, i}\,J_\nu\,{(\nu-\nu_{\rm th})\over\nu}\,{\rm d}\nu,
\label{eq:heatingt}
\end{equation}
where $n_i$ is the number density of the dissociated species.

The heating of a primordial gas comes primarily from photo-ionisation of
\h, \he~and \heplus, but there is also a small contribution from
photo-dissociation of \htwo.

\end{document}